%% file: CSP.tex
\documentclass[10pt,conference]{IEEEtran}
\IEEEoverridecommandlockouts
\usepackage[table,xcdraw]{xcolor}

\usepackage{amsmath,amssymb,amsfonts}
\usepackage{algorithmic}
\usepackage{graphicx}
\usepackage{textcomp}
\usepackage[hidelinks]{hyperref}
\def\BibTeX{{\rm B\kern-.05em{\sc i\kern-.025em b}\kern-.08em
    T\kern-.1667em\lower.7ex\hbox{E}\kern-.125emX}}
    
\include{MacrosChris}

\usepackage{mathtools}
\usepackage{cancel}
\usepackage{tabularx}
\usepackage{longtable}
\usepackage{multirow}
\usepackage{makecell}
\usepackage{lscape}
\usepackage[shortlabels]{enumitem}

\graphicspath{{Figures/}}

\usepackage{changepage}

\usepackage{tikz}
\usetikzlibrary{shapes,arrows}

\usepackage{xspace} 
\usepackage{dblfloatfix} 

\DeclarePairedDelimiter\floor{\lfloor}{\rfloor}

\makeatletter
\newcommand{\linebreakand}{%
  \end{@IEEEauthorhalign}
  \hfill\mbox{}\par
  \mbox{}\hfill\begin{@IEEEauthorhalign}
}
\makeatother

\usepackage{listings}
\colorlet{punct}{red!60!black}
\definecolor{background}{HTML}{EEEEEE}
\definecolor{delim}{RGB}{20,105,176}
\definecolor{cerulean}{rgb}{0.0, 0.48, 0.65}
\definecolor{cobalt}{rgb}{0.0, 0.28, 0.67}
\colorlet{numb}{magenta!60!black}

\lstdefinelanguage{json}{
basicstyle=\small\ttfamily,
    numbers=left,
    numberstyle=\scriptsize,
    stepnumber=1,
    numbersep=8pt,
    showstringspaces=false,
    breaklines=true,
    frame=lines,
    backgroundcolor=\color{background},
    literate=
     *{:}{{{\color{punct}{:}}}}{1}
      {,}{{{\color{punct}{,}}}}{1}
      {\{}{{{\color{delim}{\{}}}}{1}
      {\}}{{{\color{delim}{\}}}}}{1}
      {[}{{{\color{delim}{[}}}}{1}
      {]}{{{\color{delim}{]}}}}{1},
}

\lstdefinelanguage{text}{
    basicstyle= \small\ttfamily, 
    numbers=left,
    numberstyle=\scriptsize,
    stepnumber=1,
    numbersep=8pt,
    showstringspaces=false,
    breaklines=true,
    frame=lines,
    backgroundcolor=\color{background},
    columns=fullflexible,
    flexiblecolumns=true
}

\setboolean{showedits}{false}

\begin{document}

\title{Bug or not bug? That is the question}

\author{
\IEEEauthorblockN{Quentin Perez, Pierre-Antoine Jean, Christelle Urtado and Sylvain Vauttier}\\
\IEEEauthorblockA{EuroMov Digital Health in Motion, Univ Montpellier, IMT Mines Ales, Ales, France\\
\texttt{\small{Email: \{Quentin.Perez, Pierre-Antoine.Jean, Christelle.Urtado, Sylvain.Vauttier\}@mines-ales.fr}}}
}

\maketitle

\begin{abstract}
Nowadays, development teams often rely on tools such as Jira or Bugzilla to manage backlogs of issues to be solved to develop or maintain software. Although they relate to many different concerns (\eg bug fixing, new feature development, architecture refactoring), few means are proposed to identify and classify these different kinds of issues, except for non mandatory labels that can be manually associated to them.  This may lead to a lack of issue classification or to issue misclassification that may impact automatic issue management (planning, assignment) or issue-derived metrics. Automatic issue classification thus is a relevant topic for assisting backlog management. This paper proposes a binary classification solution for discriminating bug from non bug issues. This solution combines natural language processing (TF-IDF) and classification (multi-layer perceptron) techniques, selected after comparing commonly used solutions to classify issues. Moreover, hyper-parameters of the neural network are optimized using a genetic algorithm. The obtained results, as compared to existing works on a commonly used benchmark, show significant improvements on the F1 measure for all datasets.
\end{abstract}

\input{Sections/Introduction.tex}
\input{Sections/Background_v2.tex}
\input{Sections/Approach_v2.tex}

\input{Sections/Results.tex}
\input{Sections/Threats_to_validity.tex}
\input{Sections/Related_works.tex}
\input{Sections/Conclusion}

\bibliographystyle{IEEEtran}
\bibliography{IEEEabrv,bibfile}


\end{document}

%% file: MacrosChris.tex
\newcommand{\ie}{\textit{i.e.},~}
\newcommand{\eg}{\textit{e.g.},~}

\newcommand{\etal}{\textit{et al.}\xspace}
\newcommand{\etc}{etc.\xspace}

\newcommand{\github}{GitHub\xspace}

\newcommand{\secref}[1]{Section~\ref{#1}\xspace}
\newcommand{\figref}[1]{Fi\-gu\-re~\ref{#1}}

\newcommand{\tabref}[1]{Table~\ref{#1}}

\newcommand{\coderef}[1]{Listing~\ref{#1}}

\usepackage{ifthen}
\usepackage[normalem]{ulem} 
\usepackage{xcolor}

\usepackage{amssymb}
\newcommand{\nb}[3]{ \fbox{\bfseries\sffamily\scriptsize\textcolor{#2}{#1}}
{\sf\small$\blacktriangleright$\textcolor{#2}{\emph{#3}}$\blacktriangleleft$}}

\newboolean{showedits}
\setboolean{showedits}{true} 
\ifthenelse{\boolean{showedits}}
{	
\newcommand{\del}[1]{\textcolor{red}{\sout{#1}}} 
\newcommand{\paj}[1]{\nb{Pierre-Antoine}{cyan}{#1}}
\newcommand{\cu}[1]{\nb{Chris}{magenta}{#1}}
\newcommand{\sv}[1]{\nb{Sylvain}{olive}{#1}}
\newcommand{\qp}[1]{\nb{Quentin}{purple}{#1}}

\definecolor{mediumpersianblue}{rgb}{0.0, 0.4, 0.65}
\definecolor{mountbattenpink}{rgb}{0.6, 0.48, 0.55}

\newcommand\nok[1]{\textcolor{magenta}{#1}} 
}
{	
\newcommand{\del}[1]{} 
\newcommand{\gr}[1]{ }
\newcommand{\paj}[1]{ }
\newcommand{\cu}[1]{ }
\newcommand{\sv}[1]{ }
\newcommand{\qp}[1]{ }

\newcommand\nok[1]{}

}

\font\msx=msam10
\font\msy=msbm10

\newfam\msxfam \textfont\msxfam=\msx
\newfam\msyfam \textfont\msyfam=\msy

\def\famletter#1{\ifcase #1 0\or 1\or 2\or 3\or 4\or 5\or 6\or 7\or
	8\or 9\or A\or B\or C\or D\or E\or F\fi}

\edef\fx{\famletter\msxfam}

\def\@myop#1{\mathop{\mathstrut{#1}}\nolimits}

\def\_{\leavevmode \vbox{\hrule width0.5em}}

\let\xforall=\forall
\let\xexists=\exists
\let\xlambda=\lambda

\let\mc=\mathchardef

\def	\po1		{\mbox{${\cal P}_1$}}

\def	\lambda		{\@myop{\xlambda}}

\def	\forall		{\@myop{\xforall}}
\def	\exists		{\@myop{\xexists}}

\def	\dom		{\@myop{\sf dom}}
\def	\ran		{\@myop{\sf ran}}
\def	\id		{\@myop{\sf id}}
\def	\comp		{\mathbin{\raise
			0.6ex\hbox{\oalign{\hfil$\scriptscriptstyle
			\rm o$\hfil\cr\hfil$\scriptscriptstyle\rm 9$\hfil}}}}

\mc	\dres		"2\fx43
\mc	\rres		"2\fx42

\mc	\inj		"3\fx1A

\mc	\surj		"3\fx10

\def	\na1		{\mbox{${\cal N}_1$}}

\def	\int		{\mbox{${\cal Z}$}}

\def	\finse1		{\mbox{${\cal F}_1$}}

\def	\seq		{\@myop{\rm seq}}
\def	\cat		{\mathbin{\raise 0.8ex\hbox{$\mathchar"2\fx61$}}}
\def	\sep		{\hspace*{.05in}}

%% file: Sections/Introduction.tex
\section{Introduction}
\label{sec:introduction}

Bug reporting and fixing are prominent software development activities. A study conducted by Sneed \cite{1357810} evaluates that maintenance activities account for an average 20\% of the effort made by a developer on a software project. In the same way, another study conducted by the University of Cambridge \cite{britton2013reversible} estimates that the annual cost of software bugs is about \$156 billion and amounts to 50\% of maintenance costs. 
This situation creates a practical issue which is reported by Anvik \etal \cite{anvikHM05} as quotes from a Mozilla developer: "Everyday, almost 300 bugs appear that need triaging. This is far too much for only the Mozilla programmers to handle". Automated solutions are therefore needed to help developers cope with the mass of tickets that may be submitted to issue tracker tools like Jira\footnote{https://www.atlassian.com/software/jira/bug-tracking} or Bugzilla\footnote{https://www.bugzilla.org/}. Automatic issue classification, that is required for subsequent automatic management steps like prioritization, planning or assignment, still is a challenge. Detecting misclassified issues (\ie issues tagged as bugs by developers but being actually requests for enhancement) is not simpler. Systems based on textual pattern matching are not adequate because tickets are written in free form natural language. For instance, basic buzzwords detection (\eg bug) is not robust enough. Efficient classification requires sophisticated information retrieval and machine learning techniques.
This paper presents a solution based on a binary classification that separates bug and non-bug related issues. We studied and implemented a process that builds an efficient classification scheme. It relies on a combination of natural language processing (TF-IDF), statistical feature selection (Chi-square) and automatic classification (Multi-Layer Perceptron). To select this latter approach, several classification methods were benchmarked and Multi-Layer Perceptron (MLP) was selected as the most efficient. Furthermore, a MLP hyper-parameter (number of layers, layer sizes) optimization step fine-tunes the process thanks to a genetic algorithm.
Obtained results have been compared to state-of-the-art thanks to a commonly used benchmark provided by Herzig \etal \cite{herzig2013s} containing 5,591 issue tickets already labeled as bugs or non bugs. Our solution improves classification performance, evaluated by the F1 measure, on all datasets.

\noindent The remainder of this paper is organized as follows. \secref{sec:Background} presents existing binary bug classification approaches our proposal can be compared to. \secref{sec:approach} details our designed approach to create a binary bug classifier. \secref{sec:results} analyzes the obtained results. \secref{sec:threats} discusses threats to validity. \secref{sec:Related_works} presents related works on closely related problems that inspired the solution we propose in this paper. \secref{sec:conclusion} concludes and proposes perspectives for this work. 

%% file: Sections/Background_v2.tex
\section{Existing Binary Bug Classification Approaches}
\label{sec:Background}


Bug management has always been a major concern for software quality. Recent works \cite{kukkar2019novel,LimsetthoHM14,murphy2004automatic,mani2019deeptriage} have studied how to predict different bug characteristics such as their number, locations, density, or severity. Prediction systems are based on data collected from bug repositories. Various statistical models have been implemented to reach specific aims such as ticket classification, bug assignment, bug severity evaluation and duplicate bug ticket detection.

\noindent Binary bug classification, that distinguishes bug tickets from other issues, is an actively studied research subject with many practical and industrial applications. Indeed, a manual analysis of 7,000 issue tickets  from five open-source projects (HTTPClient, Jackrabbit, Lucene-Java, Rhino and Tomcat 5) concludes that 33.8\% of bug tickets were misclassified \cite{herzig2013s}.  This misclassification strongly impacts the efficiency of bug management (planing, assignment, metrics, \etc) in software development projects.
 
 \begin{table*}[t]
\centering
\caption{Herzig \etal \cite{herzig2013s} dataset details}
\label{tab:dataset-details}
\begin{tabular}{|l|c|c|c|c|c|}
\hline
\textbf{}            & \textbf{Maintainer} & \textbf{Tracker type} & \textbf{\#Reports} & \textbf{\begin{tabular}[c]{@{}c@{}}\#Labelled \\ "BUG"\end{tabular}} & \textbf{\begin{tabular}[c]{@{}c@{}}\#Labelled \\ "NBUG"\end{tabular}} \\ \hline
\textbf{HTTPClient}  & Apache              & JIRA                  & 746                & 305                                                                  & 441                                                                   \\ \hline
\textbf{Jackrabbit}  & Apache              & JIRA                  & 2443               & 697                                                                  & 1746                                                                  \\ \hline
\textbf{Lucene-Java} & Apache              & JIRA                  & 2402               & 938                                                                  & 1464                                                                  \\ \hline
\multicolumn{3}{|r|}{\textbf{Total:}}                              & \textbf{ 5591}      & \textbf{1940}                                                        & \textbf{3651}                                                         \\ \hline
\end{tabular}
\end{table*}

 \noindent After a manual curating phase, Herzig \etal \cite{herzig2013s} have labelled their dataset with tags corresponding to six categories, including a specific category for bug tickets. This provides us with a mean to easily separate bug tickets from non bug issues.
 
 \noindent \tabref{tab:dataset-details} provides details on the  dataset. Bug tickets are labelled as \texttt{BUG} while other issue tickets are labelled as \texttt{NBUG}. Many works \cite{terdchanakul2017bug,chawla2015automated,pingclasai2013classifying,luaphol2019feature,pandey2017automated,qin2018classifying} use this dataset, and more precisely a subset composed of the HTTPClient, Jackrabbit and Lucene-Java projects, as a benchmark to design and validate bug ticket automatic classification approaches. All the issue tickets extracted by Herzig \etal for these three projects come from the Bugzilla Issue Tracking System (ITS).

\begin{table*}[t]
\centering
\caption{State-of-the-art on bug classification with the dataset of Herzig \etal}
\label{tab:stateofart-classification}
\begin{tabular}{|l|c|c|c|c|c|c|} 
\hline
\multirow{2}{*}{\textbf{Study}} & \multicolumn{5}{c|}{\textbf{F1 Measure}}                                                                                                                                            & \multicolumn{1}{l|}{\multirow{2}{*}{\textbf{Evaluation Protocol}}}              \\ 
\cline{2-6}
                       & \multicolumn{1}{l|}{\textbf{HTTPClient}} & \multicolumn{1}{l|}{\textbf{Jackrabbit}} & \multicolumn{1}{l|}{\textbf{Lucence}} & \multicolumn{1}{l|}{\textbf{Mean Projects}} & \multicolumn{1}{l|}{\textbf{Cross-Project}} & \multicolumn{1}{l|}{}                                                  \\ 
\hline
Terdchanakul \etal \cite{terdchanakul2017bug} & 0.814                           & 0.805                           & 0.884                        & 0.834                              & 0.814                              & 10-fold                                                                \\ 
\hline
Chawla \etal \cite{chawla2015automated} & 0.830                           & 0.780                           & 0.840                        & 0.817                              & --                            & \begin{tabular}[c]{@{}c@{}}Training/Test split \\ 80/20 \end{tabular}  \\ 
\hline
Pingclasai \etal \cite{pingclasai2013classifying} & 0.758                           & 0.767                           & 0.818                        & 0.781                              & --                            & 10-fold                                                                \\ 
\hline
Luaphol \etal \cite{luaphol2019feature} & --                         & --                         & --                      & --                            & 0.770                              & \begin{tabular}[c]{@{}c@{}}Training/Test split \\ 70/30 \end{tabular}  \\ 
\hline
Pandey \etal \cite{pandey2017automated} & 0.687                           & 0.759                           & 0.708                        & 0.718                              & --                            & 10-fold                                                                \\ 
\hline
Qin \etal \cite{qin2018classifying} & 0.757                           & 0.771                           & 0.717                        & 0.748                              & 0.746                              & \begin{tabular}[c]{@{}c@{}}Training/Test split\\ 90/10 \end{tabular}   \\
\hline
\end{tabular}
\end{table*}

\noindent The approach proposed in this paper is evaluated on the same project subset. Our results can thus be compared to six other recent (from 2013 to 2019) works that use various ways for bug binary classification, most of them based on machine learning techniques. Their results are presented in \tabref{tab:stateofart-classification}\footnote{Confidence intervals are not reported because these figures were not provided in the papers.}. Columns \textbf{HTTPClient}, \textbf{Jackrabbit} and \textbf{Lucene} provide the F1 Measure on each subset of tickets from each project evaluated independently. The \textbf{Mean Projects} column shows the mean of those three F1 measures whereas the \textbf{Cross-Project} column gives the F1 measure calculated on the whole subset by mixing data from all three projects. 

\noindent Terdchanakul \etal \cite{terdchanakul2017bug} describe a topic-based model with N-gram IDF. N-gram IDF is used to extract key terms of any length from texts. These key terms are then used as features to classify bug reports. The classification algorithms they experiment are Logistic Regression (LR) and Random Forest (RF).
Chawla \etal \cite{chawla2015automated} describe an automated approach based on fuzzy set theory.
Pingclasai \etal \cite{pingclasai2013classifying} adopt a topic-based classification using three classifier algorithms: Decision Tree (DT), Naive Bayes (NB) and LR. Topics are modeled thanks to Latent Dirichlet Allocation. The output of this process is a collection of topic membership vectors used as input features in classifiers. F1 measure and k-fold (k=10) are used as the evaluation protocol.
Luaphol \etal \cite{luaphol2019feature} aim to discover the most efficient features for binary bug report classification. This study compares seven ways of processing textual information using sub-sequences of words: unigrams, bigrams, camel case, unigrams and bigrams, unigrams and camel case, bigrams and camel case, and all kinds of sub-sequences together. The experimental results show that unigrams may be the most efficient features for binary bug report classification. These features are processed through various classification algorithms: NB, LR, Support Vector Machine (SVM) and Radial Basis Function (RBF) kernel.
Pandey \etal \cite{pandey2017automated} and Pingclasai \etal \cite{pingclasai2013classifying} analyze how machine learning techniques may be used to perform issue classification. Authors evaluate the performance (in terms of F1 measure and average accuracy) of several classification algorithms: NB, Linear Discriminant Analysis, k-Nearest Neighbors (kNN), SVM with various kernels, Decision Trees  and RF separately. 
Qin \etal \cite{qin2018classifying} propose a bug classification method based on Long Short-Term Memory (LSTM), a typical recurrent neural network which is widely used in text classification tasks, with a softmax layer to classify data.
A softmax layer assigns decimal probabilities to each class of a multi-class problem. The sum of these decimal probabilities must be 1. This additional constraint is known to fasten learning convergence.

\noindent \tabref{tab:stateofart-classification} presents results obtained by the five works described above. The results are ordered by descending values of \textbf{Cross-Project} measure. Unfortunately, three studies \cite{pandey2017automated, chawla2015automated,pingclasai2013classifying} do not provide this information: this is why the only way to compare these results is to compute the \textbf{Mean Projects} value. Evaluation protocols are also different depending on the studies. Only three studies \cite{terdchanakul2017bug,pingclasai2013classifying,pandey2017automated} use a robust evaluation protocol, \ie k-fold cross-validation (k=10). Other works \cite{chawla2015automated,luaphol2019feature,qin2018classifying} use a less robust protocol (90/10 or 70/30 split). The lowest  \textbf{Cross-Project} measure is obtained by Qin \etal \cite{qin2018classifying} (0.746) using a Training/Test split. 

\noindent Best results are those of Terdchanakul \etal \cite{terdchanakul2017bug} on all measures. Moreover, they use a robust evaluation protocol  making them our chosen reference to compare our results to in the remaining of this paper.

%% file: Sections/Approach_v2.tex
\section{Proposed Bug Classification Approach}
\label{sec:approach}

\subsection{Approach Overview}

\begin{figure}[h]
    \centering
    \includegraphics[width=0.60\linewidth]{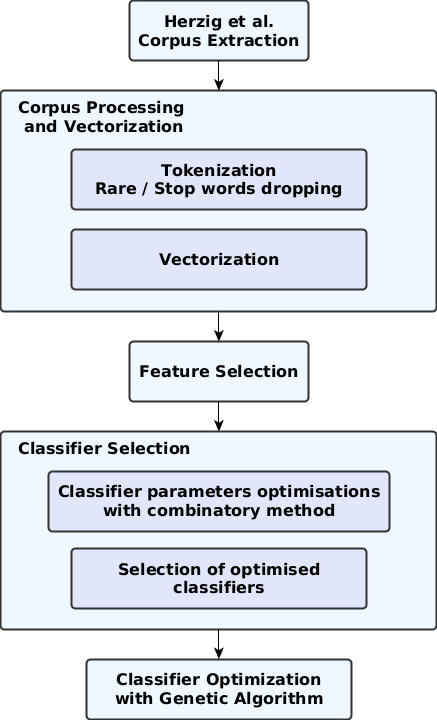}
    \caption{Bug Classification Process Overview.}
    \label{fig:process-overview2}
\end{figure}

This section provides an overview of our proposed approach for binary bug classification, from corpus extraction to classifier optimization using genetic algorithms. 
%
%
For reproducibility purposes, the source code and data used for this paper are available online\footnote{\url{https://github.com/qperez/ATTIC}}.
As shown on \figref{fig:process-overview2}, our bug classification process is composed of five main steps.

\noindent The first step is \textbf{corpus extraction}. Bug tickets all have a unique identifier as provided by the dataset of Herzig \etal \cite{herzig2013s}. These identifiers are used to retrieve the ticket corpus we use as our benchmark, using the HTTP APIs provided by the Jira ITS. Data is compiled in a unique JSON file. 

\noindent The second step is  \textbf{corpus processing}. Text is filtered (to keep only relevant information) and converted into numerical data using Natural Language Processing (NLP) techniques. 
%
%
The text of each ticket is tokenized (\ie decomposed in its smallest units
). Tuples of successive tokens (sliding windows), called n-grams, are then built. 
N-gram frequencies are calculated and n-grams are filtered when they are over-represented in the corpus. These  over-represented n-grams often correspond to stop-words (\eg articles such as "the", "an", "a"). Rare words (\ie words appearing only once or twice in a given document) are also dropped from the corpus. Finally, each ticket is represented as a vector representing the n-gram frequencies of its content. This way, each ticket is transformed into a numerical vector that provides a statistical model of its textual content in a format that can further be processed by classifiers.

\noindent The third step is  \textbf{feature selection}. Vectors representing n-grams (features) have a huge number of dimensions. However, all n-gram frequencies do not have the same importance for ticket classification. A statistical selection is therefore performed to select the best relevant dimensions in order to reduce the size of the vectors. This improves the performance of machine learning (classifier training), regarding both computation complexity and classification accuracy.

\noindent The fourth step consists in \textbf{classifier selection}. As multiple classifier types are available, selecting the right classifier type according to its performance is our first goal. Having an annotated dataset, we explored supervised learning techniques. The Scikit-learn API \cite{sklearn_api} provides six kinds of classifiers corresponding to the state-of-the-art: 

\begin{itemize}
    \item \textbf{Stochastic Gradient Descent (SGD)} uses the iterative gradient descent method to minimize an objective function written as a sum of functions: 
        \begin{displaymath}
            Q(w)=\dfrac{1}{n}\sum_{i=1}^{n}Q_i(w),
        \end{displaymath}
    where $w$ is the parameter to be estimated in order to minimize function $Q(w)$. $Q_i$ corresponds to the $i\text{-th}$ observation in the training dataset.
    
    \item \textbf{Support Vector Machines (SVM)} are non probabilistic classifiers based on linear algebra. Training SVMs creates hyperplanes that separate multidimensional data into different classes. SVMs optimize hyperplane positions by maximizing their distance with the nearest data. These classifiers generally reach a good accuracy. 
    
    \item \textbf{Random Forest (RF)} is a parallel learning method based on multiple randomly constructed decision trees. Each tree of the random forest is trained on a random subset of data according to the bagging principle, with a random subset of features according to the principle of random projections.
    
    \item \textbf{Ridge Regression (RR)} is based on linear least squares regression with a $\mathit{L2}$ regularization (known as Tikhonov's): a penalty equal to the sum of the squares of the weights, multiplied by a $\alpha$ penalty strength factor, is added to the $f_{loss}$ objective function being minimized. Ridge regression can thus be defined as follows:
        \begin{displaymath}
            Obj = f_{loss}(w) + \alpha \cdot \sum{}^{} w^2
        \end{displaymath}
     $f_{loss}$  depends on the underling task (\eg cross-entropy loss for classification). $\alpha$ is generally adjusted during model validation and is called the regularization parameter.
    
    \item \textbf{k-Nearest Neighbors (kNN)} is a non-parametric method in which the model stores the data of the training dataset to perform the the classification of the test dataset. To assess the class of a new input, kNN looks for its closest k neighbors using a distance formula (\eg Euclidean distance) and chooses the class of the majority of neighbours.
    
    \item \textbf{Multi-Layer Perceptron (MLP)} is a type of formal neural network that is organized in several layers. Information flows from the neurons of the input layer to the neurons of the output layer through weighted connections. Supervised training incrementally adjusts the weights of connections (error backpropagation) so that the expected outputs can be learned by the MLP. Through the use of multi-layers, a MLP is able to classify data that is not linearly separable (using multiple learned hyperplanes).
\end{itemize}

\noindent Main classifier parameters, called their hyper-parameters, are tested by a combinatorial method to find the best configuration. Finally, the classifier that produces  best results is selected: MLP performs best and will thus be used in the remaining.

\noindent The last step is MLP \textbf{classifier optimization using a genetic algorithm}. Optimization regards number of input features (\ie size of the input layer) and the structure of the MLP (number and sizes of hidden layers). We chose these parameters because their optimization is not efficient with combinatorial or dichotomous methods (as compared to the hyper-parameters configured in the previous step). 

\noindent The result of our approach is a classifier based on MLP with its  hyper-parameters optimized with a genetic algorithm. Performances of our optimized MLP classifier will further be compared to results obtained by Terdchanakul \etal \cite{terdchanakul2017bug} in \secref{sec:results}.

\noindent After this coarse grained presentation, next subsection provides further details on our approach.

\subsection{Detailed Approach}

\subsubsection{Bug Ticket Corpus Extraction}
Data used in this study are based on the corpus of Herzig \etal \cite{herzig2013s}. 5,991 bug tickets are extracted from three popular software projects: \textit{Lucene}, \textit{JackRabbit} and \textit{HttpClient}. 
After the raw extraction, tickets are mapped to their classification (\ie bug or not bug) as given by the Herzig corpus, thanks to corresponding ticket identifiers. Finally, a JSON file containing the set of tickets and their classification is produced. This file is subsequently used for text processing but also to test and train classifiers. A sample of annotated issue ticket is given by \coderef{lst:sample-ticket-extraction}. The JSON data structure is composed of six fields: 
\begin{itemize}
    \item \texttt{key} represents the unique identifier for the issue ticket in the ITS, named ticket ID.
    \item \texttt{summary} is the short description of the issue.
    \item \texttt{description} is the long description of the issue. For a bug, it should describe the way it appears and the problems it causes.
    \item \texttt{classification} is the category in which Herzig \etal classified the issue.
    \item \texttt{type} is the issue category as mentioned by the issue opener in the repository.
    \item \texttt{label} is our binary classification of the issue: \texttt{BUG} for a ticket classified by Herzig \etal as a bug (in field \texttt{classification}). Other categories arc classified \texttt{NBUG}.
\end{itemize}

\begin{lstlisting}[caption={Sample issue ticket used to train and test classifiers.},captionpos=b,float,label={lst:sample-ticket-extraction},language=json,numbers=none]
{
  "key": "HTTPCLIENT-126",
  "summary": "Default charset",
  "description": "As defined in RFC2616 the default character set is ISO-8859-1 an not US-ASCII \nas defined in HttpMethodBase. See \"3.7.1 Canonicalization and Text Defaults\" at\nRFC 2616",
  "classified": "IMPROVEMENT",
  "type": "BUG",
  "label": "NBUG"
}
\end{lstlisting}

\subsubsection{Corpus Processing}
\label{subsec:corpus_processing}

\begin{figure*}[h]
    \centering
    \includegraphics[width=0.92\linewidth]{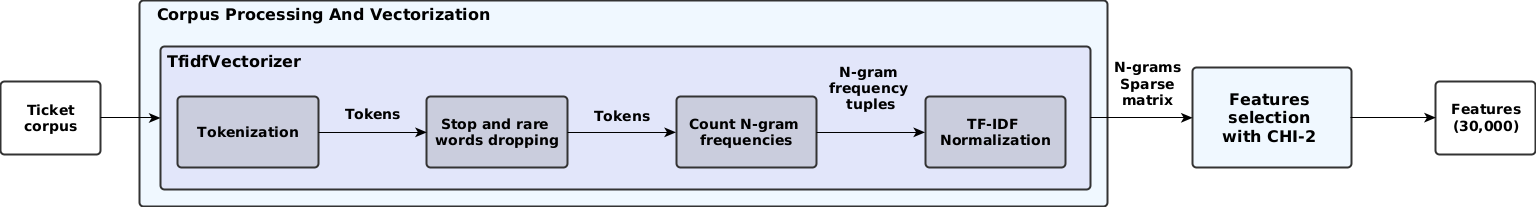}
    \caption{Corpus processing steps.}
    \label{fig:corpus-processing}
\end{figure*}

After the extraction, a bag-of-word processing is performed on the corpus in order to code textual data into a vector representation that classifiers can use. This step is not neutral as it may significantly impact the classifier performance. We rely on the Scikit-learn API \cite{sklearn_api}, which is widely used in machine learning projects conducted by public laboratories and companies as a standard, configurable framework providing a wide range of methods. As shown in \figref{fig:corpus-processing}, corpus processing is divided into two main steps: Natural Language Processing and Feature Selection.
\newline\newline
\textbf{Natural Language Processing}. In this step, the corpus is cleaned and then projected into a vector representation. \texttt{TfidfVectorizer} provided by the Scikit-learn API \cite{sklearn_api} is used to process the textual data and compute statistical data using the Term Frequency Inverse Document Frequency (TF-IDF) method \cite{jones1972statistical}. 
    \begin{enumerate}[(a)]
        \item \textbf{Tokenisation}. Textual data is divided into text units called tokens. Each token $tok$ represents a word. From these tokens, n-grams are created. A n-gram is a sequence of $n$ contiguous tokens that can formally be written as follows: $n\text{-gram} = \{tok_1,tok_2,...,tok_n\}$. In this paper we use uni-grams, bi-grams and tri-grams. \coderef{lst:sample-unigrams} shows some uni-grams created from the bug ticket of \coderef{lst:sample-ticket-extraction}. In turn, \coderef{lst:sample-bigrams} and \coderef{lst:sample-trigrams} respectively show samples of bi-grams and tri-grams.

            \begin{lstlisting}[caption={Sample of uni-grams created with the last sentence in \coderef{lst:sample-ticket-extraction}.},captionpos=b,label={lst:sample-unigrams},language=text,numbers=none, xleftmargin=\dimexpr-\csname @totalleftmargin\endcsname]
"see", "371", "canonicalization", "text", 
"defaults", "rfc", "2616"
            \end{lstlisting}

            \begin{lstlisting}[caption={Sample of bi-grams created with the last sentence in \coderef{lst:sample-ticket-extraction}.},captionpos=b,label={lst:sample-bigrams},language=text,numbers=none, xleftmargin=\dimexpr-\csname @totalleftmargin\endcsname]
"see 371", "371 canonicalization"
"canonicalization text", "text defaults",
"defaults rfc", "rfc 2616"
            \end{lstlisting}
            
            \begin{lstlisting}[caption={Sample of tri-grams created with the last sentence in \coderef{lst:sample-ticket-extraction}.},captionpos=b,label={lst:sample-trigrams},language=text,numbers=none, xleftmargin=\dimexpr-\csname @totalleftmargin\endcsname]
"see 371 canonicalization", 
"371 canonicalization text",
"canonicalization text defaults",
"text defaults rfc", "defaults rfc 2616"
            \end{lstlisting}
       
        \item \textbf{Stop words and rare words dropping}.
        Tokens with high occurrence rates in the corpus (often stop words) are not relevant for classification. Hence, tokens appearing in more than 50\% of the tickets of the corpus are filtered.
        Conversely, tokens with very low occurrence rates are also often not relevant. Thus, tokens appearing in less than 2 tickets from the corpus are also filtered.
        An issue ticket contains two main parts: a summary and a description. An example of summary and description is given on  \coderef{lst:sample-ticket-extraction}. The summary part is actually the title of the ticket while the description part explains the reported issue. Some studies \cite{pandey2017automated} use only the summary and ignore the description. In our work, we have experimented classification using only the summary or only the description. However, results are better when both are used. Besides, to capitalize on previous studies \cite{pandey2017automated,ko2006linguistic} about the importance of the summary, we have experimented to increase its weight by duplicating its content. After iterative tests, it happens that duplicating three times the contents of the ticket summary produces the best results.
        
        \item \textbf{N-gram frequency calculation}. Occurrences for each uni-gram, bi-gram or tri-gram is counted. As a result, a tuple is produced for each n-gram containing its frequency in each ticket.
        
        \item \textbf{TF-IDF normalisation}. The Term Frequency Inverse Document Frequency (TF-IDF) method implemented in Scikit-learn is used to weight and normalize n-gram frequencies in the corpus. This statistical method evaluates the importance of a $t$ term (a n-gram) contained in a $d$ document (a ticket), relatively to the $D$ corpus of documents (the whole ticket dataset).
        
        The weight increases with the number of occurrences of the $t$ term in the $d$ document and decreases with the frequency of the $t$ term in the $D$ corpus thus emphasizing terms which presence is discriminating.
        
        The TF-IDF formula is: 
        \begin{displaymath}
            tfidf(t,d,D) = tf(t,d) \cdot idf(t,D)
        \end{displaymath}
        $tf$ corresponds to the number of occurrences of a $t$ term in the $d$ document (noted $n_{t,d}$) divided by the total number of terms in the $d$ document (noted $n_{d}$):
        \begin{displaymath}
            tf(t,d) = 1 + \log{\left(\dfrac{n_{t,d}}{\sum_{}^{}n_{d}}\right)}
        \end{displaymath}
        $tf$ is calculated as $1 + log(tf)$ to cap the weight of very frequent terms.
        
        $idf$ represents the importance of the $t$ term in the $D$ corpus as a whole. The number of documents where the $t$ term appears in the $D$ corpus is noted $n_{t,D}$: 
        \begin{displaymath}
            idf(t,D) = \log{\left(\dfrac{1 + |D|}{1 + n_{t,D}}\right) + 1}
        \end{displaymath}
        $idf$ is defined here in its smooth version by adding $1$ after the $log$ function. Constant 1 is added on numerator and denominator to prevent division by zero.
        
        This process returns a sparse matrix with TF-IDF weights for all n-grams and all tickets of the corpus.
    
    \end{enumerate}
    \vspace{\baselineskip}
    \textbf{Feature selection}. Selection of the best features (the n-grams) is performed before classifier training in order to avoid irrelevant and noisy features. Indeed, the initial number of features is huge: 24,496 features considering only uni-grams, 63,925 features counting both uni-grams and bi-grams and 99,351 features including uni-grams, bi-grams and tri-grams. The number of selected features impacts training time, model fitting and classifier quality. 
    
    \noindent A Chi-square test is used to measure the association relation between our categorical variables (\texttt{BUG} or \texttt{NBUG}) and our features. The Chi-square test statistically selects a number of most relevant features for each categorical variable that is fixed by the user.
    We empirically have fixed the number of features to be selected by the Chi-square test to 30,000. To determine this value, we have sampled the feature number in the range $[5,000-60,000]$ with 5,000 increments. For each sample value, we have computed the F1 measure of each classifier tested in this paper (using 10-fold validation). Below 30,000, all F1 measures are very low which means that classifiers all perform poorly. For each value over 30,000, F1 measures of the tested classifiers are too different to be easily comparable. It thus appears that 30,000 is the best trade-off for the global performance of classifiers. 
    

\subsubsection{Classifier Selection}
\label{subsec:classifier-selection}
Several classifiers are compared in order to select the best suited one. To do so, their hyper-parameters are set using a brute force optimization method and their performances compared on sets of 30,000 selected features.

\begin{figure*}[h]
    \centering
    \includegraphics[width=0.72\linewidth]{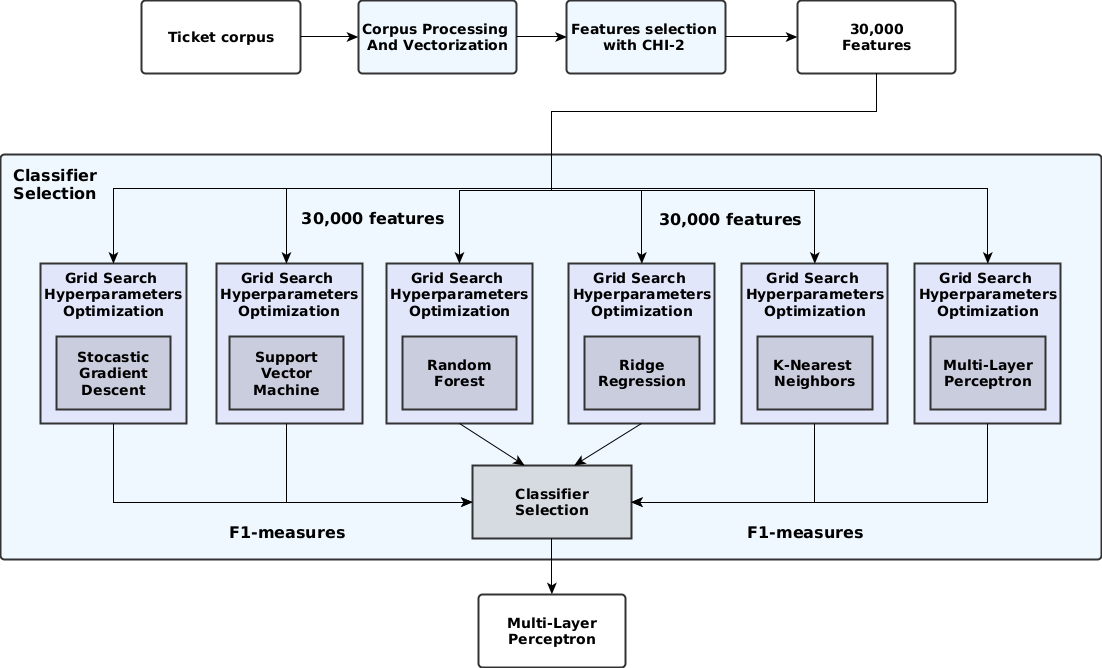}
    \caption{Classifier selection process.}
    \label{fig:selection_classifier}
\end{figure*}

 \noindent \figref{fig:selection_classifier} shows the six classifiers used in this study and described above. All are configured with hyper-parameters which values influence the model building process. Choosing correct values for hyper-parameters is crucial to have the best possible prediction. To do so, a Grid-Search algorithm is used to optimize the hyper-parameters of each classifier. Grid-Search, also called parameter sweep, is a brute-force method that searches for an optimal parameter value combination using their $n$-fold Cartesian product.  
A discrete set of values to be tested is provided for each hyper-parameter. All possible $n$-tuples of hyper-parameter values are generated and used to initialize the classifier. Finally classifiers are trained with 30.000 features and evaluated by a 10-fold. The Multi-Layer Perceptron (MLP) produces the best results (as evaluated by the F1 measure) and is therefore selected as our classification technique.

\subsubsection{Classifier Optimization}
\label{subsec:classifier-optimization-GA}

Hyper-parameter value optimization is central for prediction quality. Hyper-parameter values cannot be fine-tuned by the coarse Grid-Search  method proposed above, that requires lists of discrete values to create combinations. A precise classifier optimization needs to traverse a continuous value interval for each hyper-parameter to better explore the search space.

\noindent To do so, our MLP classifier is further optimized using a genetic algorithm (GA) \cite{10.2307/24939139}. GAs can optimize values precisely within continuous value intervals.

\noindent Three MLP hyper-parameters must be optimized: 
\begin{itemize}
    \item number of features: size of the feature vectors used to train and test by the MLP classifier (equivalent to the number of neurons on the input layer of the MLP),
    \item number of hidden layers for the MLP,
    \item hidden layers sizes: numbers of neurons on each hidden layer.
\end{itemize}

\begin{figure}[]
    \centering
    \includegraphics[width=0.9\linewidth]{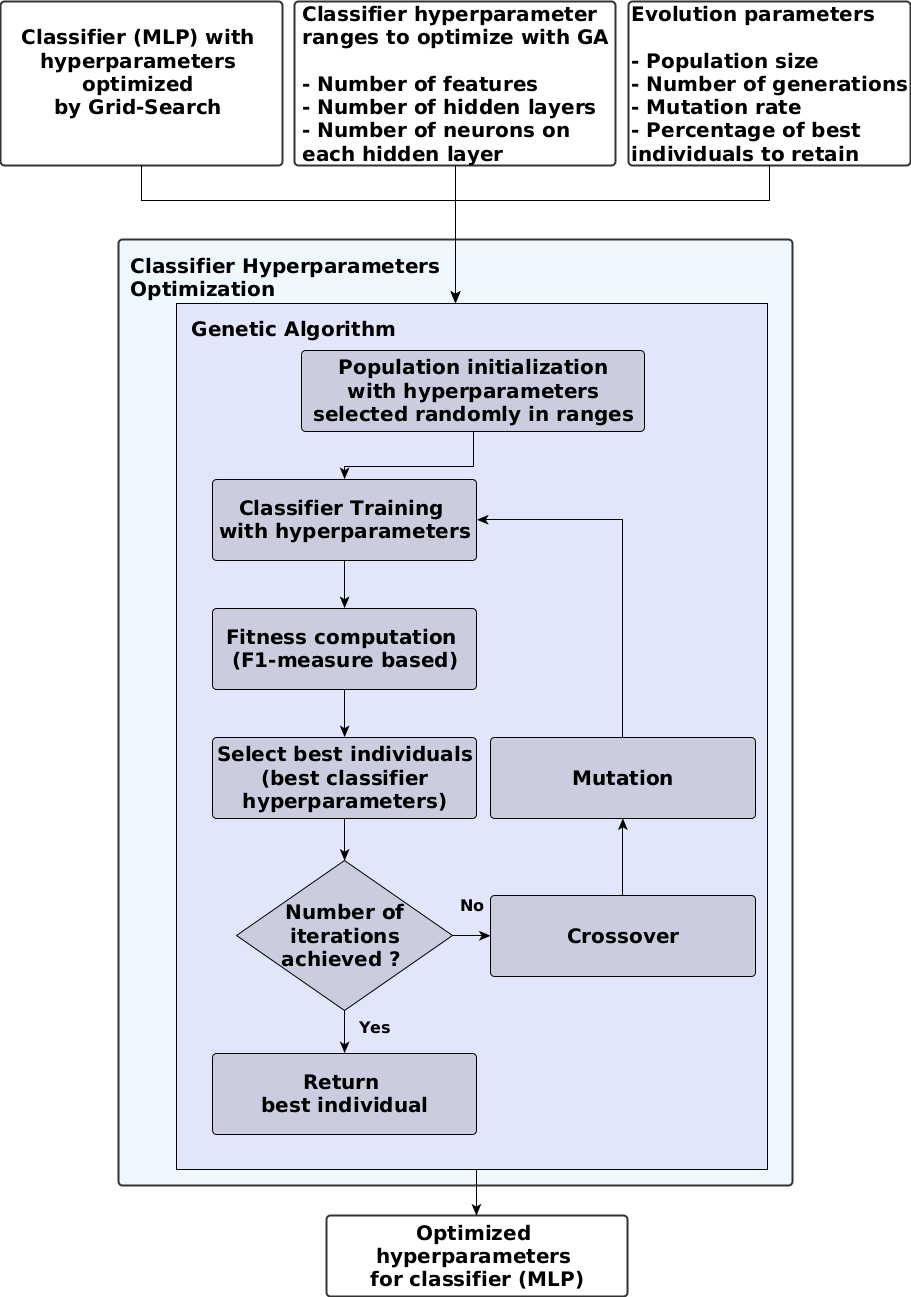}
    \caption{Evolutionary system implemented to find best MLP parameters.}
    \label{fig:evolutionary-system}
\end{figure}

\noindent \figref{fig:evolutionary-system} presents the GA we have implemented. GAs are dedicated to the search for optimal solutions using stochastic techniques inspired from evolutionary biological mechanisms such as mutation, crossover and natural selection. An initial set of solutions, called the population, is randomly generated. Each solution, called an individual, is described by a set of characteristics stored in a list of values called its chromosome. Each individual is evaluated thanks to an objective function. The score of each individual is called its fitness. Like in natural selection theory, individuals with the highest fitness have the best chances to survive and to reproduce themselves. A subset of the best individuals, called the parents, are selected to generate a set of new individuals, called their offspring. Each individual in the offspring is generated by a couple of parents thanks to a crossover of their chromosomes. Random mutations are applied to maintain diversity in the population in order to balance search space exploration with optimization convergence. 
 The offspring replace the least fitted individuals and creates a new generation of the population (which size remains constant throughout generations). This process is iterated until a stop condition is reached (for instance a time limit or a maximum number of iterations). Thanks to this iterative selection and the recombination of the best known solutions, GA are eventually able to find optimized and even optimal solutions for complex combinatorial problems.  

\noindent The initial population is generated using hyper-parameter values that are chosen randomly inside fixed intervals:
\begin{itemize}
    \item number of features ($[20,000-60,000]$). These bounds correspond to the lower bound used in the Grid-Search step and approximately two-thirds of the maximum number of features (99,351). The value empirically determined during the feature selection phase is in this interval (30,000). 
    \item number of hidden layers ($[2-15]$). The upper bound is chosen to limit computation time.
    \item hidden layer sizes ($[1-30]$). This range creates variability in the generated structures with reasonable computation time.
\end{itemize}

\noindent Besides, our proposed GA has its own hyper-parameters that control different aspects of its evolutionary strategy. These parameters are used to create a population, select the best individuals and introduce diversity at each generation:
\begin{itemize}
    \item Population size. We run different experiments with 50, 100, 200 and finally 300 individuals, following De Jong's recommendation to chose moderated population sizes \cite{JongS89}.
    \item Number of generations. It is fixed to 150 iterations to have an acceptable computing time. 
    \item Percentage of best individuals to retain in the population ($p_{ret} = 20\%$). 
    \item Probability of mutation for each individual ($p_{mut} = 0.1$). The $p_{mut}$ value has been studied in many works \cite{JongS89,razali2011genetic,Gen1997} and is thus fixed to its recommended value.
    \item Random selection level ($p_{sel} = 0.3$). This parameter controls the probability to randomly select a parent in order to maintain diversity in the population \cite{thierens349898}. 
\end{itemize}

\noindent As shown in \figref{fig:evolutionary-system}, the GA processes seven steps:

\begin{enumerate}
    \item The population is initialized with randomly generated individuals. Each individual is described by hyper-parameter values used to configure its corresponding MLP: a number of features and a tuple of hidden neuron layer sizes (tuple size corresponds to the number of hidden layers). An individual $i$ can be described for instance as $i(25540, (11,25,18,6,13))$: 25540 features (or entry neurons in first layer) and 5 neuron layers of respectively 11, 25, 18, 6 and 13 neurons from the first hidden layer to the last hidden layer.

    \item The MLP corresponding to each new individual is trained, using a 75/25 split of train/test dataset. This strategy is chosen for its speed and low computational cost.

    \item Fitness of individuals is based on the performance of the MLP on the test dataset. F1 measure is chosen because it requires both high precision and high recall. 

    \item Best individuals are selected according to their fitness and the percentage ($p_{ret}$) of individuals to retain . 

    \item The selected individuals reproduce and their offspring is generated thanks to a crossover of their chromosomes. Couples of parents are randomly chosen among the selected individuals. Crossover is performed both on the number of features and on the neurons layers :
    \begin{itemize}
        \item Crossover on the number of features is performed using a mean between the number of features of the two parents.
        \item Crossover on neurons layers is done using a single cut point at the middle of the hidden layers of the two parents $a$ and $b$ \cite{10.1007/3-540-54563-8_104}. If the number of layers is odd the cutting point is rounded down. Children thus inherit half of their structures from each of their parents, as shown in the following example. The first half of the layers for parent $a$ is transmitted whereas the rest is deleted. For parent $b$, the second half is transmitted.  
    \end{itemize}
    \begin{gather*} 
        \textcolor{cobalt}{i_{parent_a}(21912, (12,23,8,4))}\\ Crossover \\ \textcolor{mountbattenpink}{i_{parent_b}(30023, (4,23,5,13,27))} \\
        \Downarrow \\
        i_{child}(\floor{(\textcolor{cobalt}{21912} + \textcolor{mountbattenpink}{30023}) / 2}, 
        \textcolor{cobalt}{(12,23,\bcancel{8},\bcancel{4})} \cdot \textcolor{mountbattenpink}{(\bcancel{4},\bcancel{23},5,13,27)}) \\
        \Downarrow \\
         i_{child}(25967,\textcolor{cobalt}{(12,23} \textcolor{mountbattenpink}{,5,13,27)})
    \end{gather*}
    \item Mutation is performed. It consists in changing a random gene in an individual. In this approach, mutation impacts solely hidden neuron layers. Three kinds of mutation, randomly chosen from with an equal probability ($ = \frac{1}{3}p_{mut}$ each), are possible:
    \begin{itemize}
        \item Addition: adds a neurons layer randomly generated in the individual. 
        Example: 
            \begin{gather*} 
                i(21912, (12,23,45)) \\
                Addition \\
                \Downarrow \\
                i(21912, (12,23,45,18))
            \end{gather*}
            \vspace{-.65cm}
        \item Deletion: deletes a neurons layer in the individual. Example: 
            \begin{gather*} 
                i(21912, (12,23,45)) \\
                Deletion \\
                \Downarrow \\
                i(21912, (12,45))
            \end{gather*}
            \vspace{-.65cm}
        \item Substitution: chooses a number of neurons randomly in the hidden layer sizes bounds. Then selects randomly a layer in the individual to change it by the number of neurons previously chosen. Example: 
            \begin{gather*} 
                i(21912, (12,23,45)) \\
                Substitution \\
                \Downarrow \\
                i(21912, (7,23,45))
            \end{gather*}
    \end{itemize}
    \vspace{.5cm}
    
    \item The best individual found is returned when the fixed number of generations is reached. This individual provides the best optimized values found for MLP hyper-parameters.

\end{enumerate}

%% file: Sections/Results.tex
\section{Results}
\label{sec:results}

\subsection{Classifier Selection}

As stated in \secref{subsec:classifier-selection}, we compare the performance of six kinds of classifiers using F1 measure: Stochastic Gradient Descent (SGD), Support Vector Machines (SVMs), Random Forest (RF), Ridge Regression (RR), k-Nearest Neighbors (kNN) and Multi-Layer Perceptron (MLP). Hyper-parameters for each classifier coarsely optimized by the Grid-Search step and performances are evaluated with a 10-fold. Results for this experiment are presented in \tabref{tab:baseline-results}. The Grid-Search algorithm finds a set of optimized hyper-parameters for each classifier (see \tabref{tab:baseline-results}). Using these hyper-parameters, MLP obtains the best results (0.868). Hence, MLP is selected for our classification approach and parameterized as proposed by the Grid-Search algorithm as a starting configuration.
\begin{table}[h]
\centering
\caption{Results obtained with SciKit classifiers using a 10-fold cross-validation and 30,000 selected features.}
\label{tab:baseline-results}
\begin{tabular}{|l|c|c|c|}
\hline
\textbf{Classifier}  & \textbf{\makecell{Grid-Search Optimized \\ Classifier Parameters}}                   & \textbf{F1 Measure} & \textbf{\makecell{Evaluation\\Protocol}} \\\hline
\textbf{MLP}     & \makecell{activation='tanh'\\learning\_rate='adaptive'\\max\_iter=100\\random\_state=0}          & \makecell{\textbf{0.868}\\\textbf{CI 95\%: 0.034}}  & 10-fold \\ \hline
SVM     & \makecell{C=100 \\ gamma='scale'}                                                           & \makecell{0.857\\CI 95\%: 0.042}  & \makecell{10-fold}  \\ \hline
SGD     & \makecell{loss='modified\_huber'\\ max\_iter=5000\\random\_state=0}                                      &  \makecell{0.841\\CI 95\%: 0.037}  & \makecell{10-fold} \\ \hline
RR      & \makecell{random\_state=0}                                                                                &  \makecell{0.819\\CI 95\%: 0.050}  & \makecell{10-fold} \\ \hline
RF      & \makecell{criterion='entropy'\\ n\_estimators=20\\random\_state=0}                                         &  \makecell{0.610\\CI 95\%: 0.089}  & \makecell{10-fold} \\ \hline
kNN     &  \makecell{weights='distance'\\n\_neighbors=2}     &  \makecell{0.449\\CI 95\%: 0.107}  & \makecell{10-fold} \\ \hline
\end{tabular}

\end{table}


\subsection{Intermediate Results}
\label{subsec:intermediate_results}

\begin{figure*}[b]
    \centering
    \includegraphics[width=1\textwidth]{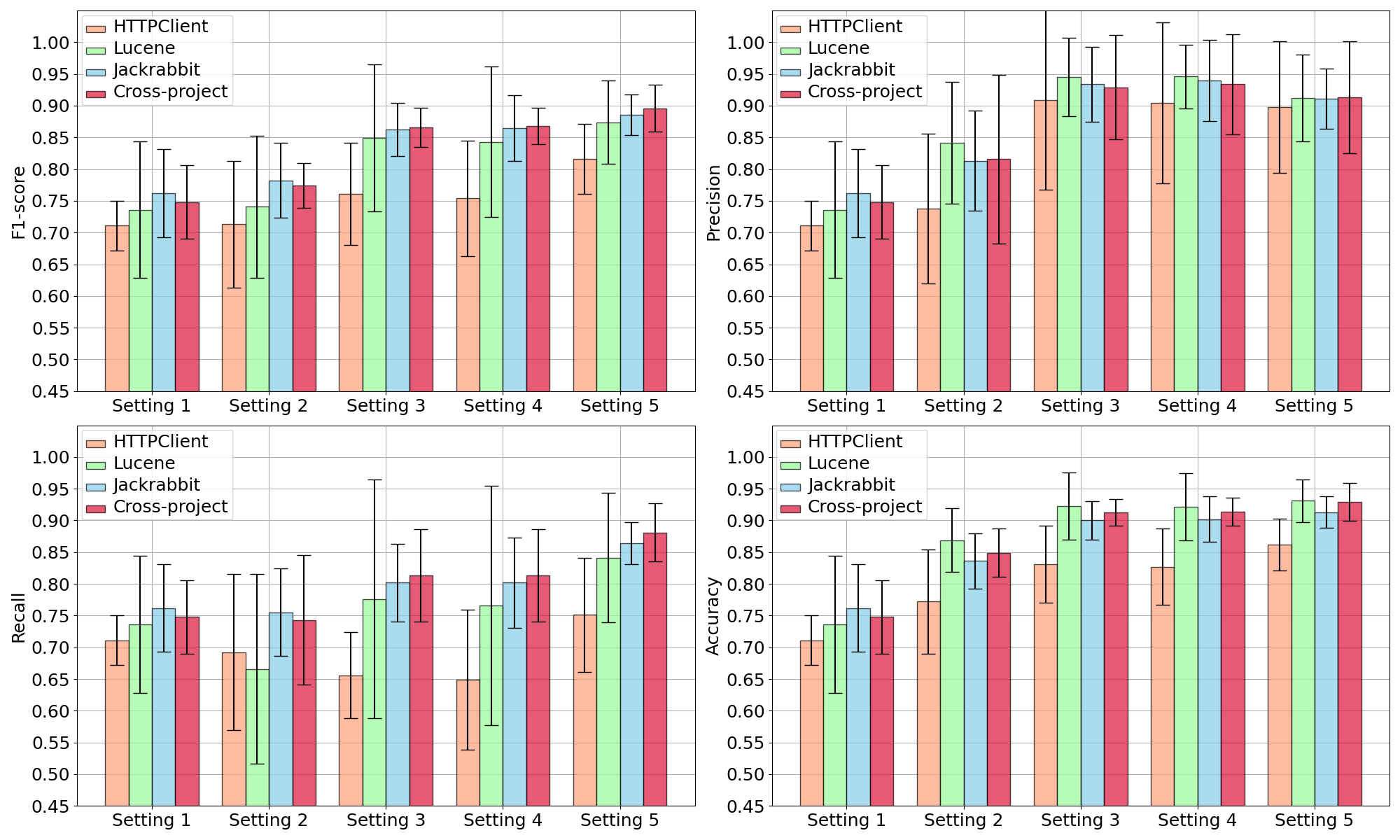} 
    \caption{Result of measurements on the 5 settings tested.}
    \label{fig:setting_results}
\end{figure*}

In order to fine-tune our classification process, we study the influence of each step on global performance. We therefore compute F1 Measure and accuracy metrics for five different settings. Results are shown in \figref{fig:setting_results}.

\noindent Setting 1 uses uni-grams and TF-IDF without logarithmic term frequency attenuation. No feature selection is done. MLP uses the default hyper-parameter values set by the Scikit-Learn API. 

\noindent In Setting 2, TF-IDF vectorization uses uni-grams, bi-grams and logarithmic term frequency attenuation. Rare and stop words are  filtered. All features are used (63,924). MLP hyper-parameters are set to default. A noticeable gain in terms of precision (+0.068) and accuracy (+0.101) can be measured on cross-project as compared to Setting 1. F1 measure improvement is smaller (+0.027) because of a lower recall.

\noindent In Setting 3, TF-IDF vectorization uses uni-grams, bi-grams and tri-grams. This setting also performs a Chi-square selection on features (30,000 features). MLP hyper-parameters are still set to their default. Tri-grams and feature selection have a positive influence on the F1 measure (+0.092) thanks to improvements on recall (+0.070) as compared to Setting 2.

\noindent Setting 4 uses the same TF-IDF vectorization and feature selection as in Setting 3. The MLP is configured with the hyper-parameter values optimized by the Grid-Search step (see \tabref{tab:baseline-results}). Gain of performance on all measures is close to null. 

\noindent Setting 5 uses same TF-IDF and feature selection as in Settings 3 and 4. MLP hyper-parameters are now optimized by the GA. Recall is improved for all projects and on cross-project (+0.068) as compared to Setting 4. Precision is slightly degraded but remains rather stable, especially on cross-project. This way, F1 measure is improved for all projects and on cross-project (+0.031). Classification is thus a little less precise but more robust. 

\noindent Results of GA and comparison with the state-of-the-art are described in the following section. 

\subsection{Classifier Fine-Tuning with a Genetic Algorithm}
\label{subsec:ga_section_results}

As stated in \secref{subsec:classifier-optimization-GA}, 
we optimize the hyper-parameters of the MLP classifier thanks to a GA. Besides, we use the other options of Setting 4, as described in \secref{subsec:intermediate_results}.

\begin{figure}[h]
    \centering
    \includegraphics[width=0.9\linewidth]{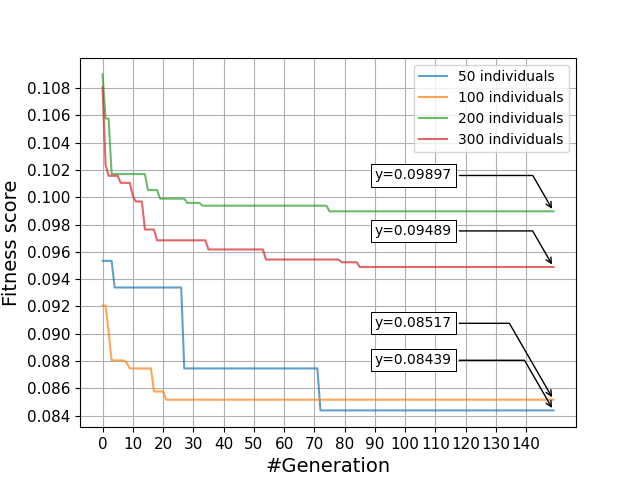}
    \caption{Fitness evolution through 150 generations with genetic algorithm.}
    \label{fig:fitness}
\end{figure}

\noindent The GA is run with hyper-parameter values as defined in \secref{subsec:classifier-optimization-GA} on 150 generations. \figref{fig:fitness} shows the evolution of fitness over generations with four different population sizes. Smaller population sizes with 50 and 100 individuals have the best results while larger populations have the worst results. These differences can be due to the parent selection mechanism we implemented. As parents are randomly selected in a population subset based on a fixed proportion of the best individuals (20\%), smaller population entails smaller parent sets, containing on average better fitted individuals. This may result in a more elitist evolution strategy leading to a quicker fitness improvement. Confirmation of this hypothesis is a perspective. 
\noindent GA executed with 50 individuals and 150 generations returns an optimized individual with a configuration that uses 7 hidden layers with sizes varying from 9 to 15 and 37,362 input features: $i(37362, (15, 9, 10, 11, 9, 15, 11))$. 

\noindent To compare the best individual returned by the GA to state-of-the-art results, we use this individual to perform a 10-fold validation on the benchmark of 5,591 tickets. Our results are presented in \tabref{tab:results_gen_algo} as compared to those obtained by Terdchanakul \etal (best results from state-of-the-art).

\begin{table*}
\centering
\caption{Comparison between results obtained by Terdchanakul \etal \cite{terdchanakul2017bug} and our approach.}
\label{tab:results_gen_algo}
\begin{tabular}{|c|c|c|c|c|c|l|c|c|c|} 
\hline
\multicolumn{9}{|c|}{\textbf{F1 Measure} }                                                                                                                                                                                                                                                                                                                                                                                                                                                                                                                                                                                                                 & \begin{tabular}[c]{@{}c@{}}\textbf{Evaluation}\\\textbf{ Protocol} \end{tabular}  \\ 
\hline \hline
\rowcolor[rgb]{0.937,0.937,0.937} \multicolumn{9}{|c|}{Terdchanakul \etal \cite{terdchanakul2017bug}}                                                                                                                                                                                                                                                                                                                                                                                                                                                                                                                                                                                       & {\cellcolor[rgb]{0.937,0.937,0.937}}                                              \\ 
\cline{1-9}
\rowcolor[rgb]{0.937,0.937,0.937} \multicolumn{2}{|c|}{HTTPClient}                                                                               & \multicolumn{2}{c|}{Jackrabbit}                                                                                & \multicolumn{2}{c|}{Lucene}                                                                                    & Mean Projects                                                                                                                      & \multicolumn{2}{c|}{Cross-Project}                                                                                               & {\cellcolor[rgb]{0.937,0.937,0.937}}                                              \\ 
\cline{1-9}
\rowcolor[rgb]{0.937,0.937,0.937} \multicolumn{2}{|c|}{0.814}                                                                                    & \multicolumn{2}{c|}{0.805}                                                                                     & \multicolumn{2}{c|}{0.884}                                                                                     & \multicolumn{1}{c|}{\textbf{0.843}}                                                                                                & \multicolumn{2}{c|}{\textbf{0.814} }                                                                                             & \multirow{-3}{*}{{\cellcolor[rgb]{0.937,0.937,0.937}}10-Fold}                     \\ 
\hline \hline
\rowcolor[rgb]{0.855,0.91,0.988} \multicolumn{9}{|c|}{Approach presented in this paper: MLP optimized with GA}                                                                                                                                                                                                                                                                                                                                                                                                                                                                                                                                                                               & {\cellcolor[rgb]{0.855,0.91,0.988}}                                               \\ 
\cline{1-9}
\rowcolor[rgb]{0.855,0.91,0.988} \multicolumn{2}{|c|}{HTTPClient}                                                                                & \multicolumn{2}{c|}{Jackrabbit}                                                                                & \multicolumn{2}{c|}{Lucene}                                                                                    & Mean Projects                                                                                                                      & \multicolumn{2}{c|}{Cross-Project}                                                                                               & {\cellcolor[rgb]{0.855,0.91,0.988}}                                               \\ 
\cline{1-9}
\rowcolor[rgb]{0.855,0.91,0.988} \begin{tabular}[c]{@{}>{\cellcolor[rgb]{0.855,0.91,0.988}}c@{}} 0.816\\ (+0.002) \end{tabular} & CI 95\%: 0.055 & \begin{tabular}[c]{@{}>{\cellcolor[rgb]{0.855,0.91,0.988}}c@{}}0.886\\ (+0.081) \end{tabular} & CI 95\%: 0.032 & \begin{tabular}[c]{@{}>{\cellcolor[rgb]{0.855,0.91,0.988}}c@{}}0.874\\ (-0.010) \end{tabular} & CI 95\%: 0.066 & \multicolumn{1}{c|}{\begin{tabular}[c]{@{}>{\cellcolor[rgb]{0.855,0.91,0.988}}c@{}}\textbf{0.859}\\\textbf{(+0.016)}\end{tabular}} & \begin{tabular}[c]{@{}>{\cellcolor[rgb]{0.855,0.91,0.988}}c@{}}\textbf{0.896}\\\textbf{ (+0.082)} \end{tabular} & CI 95\%: 0.037 & \multirow{-3}{*}{{\cellcolor[rgb]{0.855,0.91,0.988}}10-Fold}                      \\
\hline
\end{tabular}
\end{table*}

\noindent Our proposed approach obtains a gain of +0.016 on the mean F1 measure over all the projects, as compared with the scores obtained by Terdchanakul \etal. Our solution improves results on two out of three projects: HTTPClient (+0.002) and Jackrabbit (+0.081). A low deterioration is observed on Lucene (-0.010). F1 measure on cross-project is much improved (+0.081) increasing from 0.814 to almost 0.9. Our result (0.896) is also very good because it is very close to 1. As shown in \figref{fig:setting_results}, our solution achieves 0.881 for recall, 0.913 for precision and 0.929 for accuracy. \newline

\noindent Final result: \newline

\noindent \fbox{
\begin{minipage}{0.93\columnwidth} 
Our results are not only high but also well balanced between recall and precision, thus highlighting the quality of our proposed classifier, that outperforms scores obtained by Terdchanakul \etal.
\end{minipage} 
}


%% file: Sections/Threats_to_validity.tex
\section{Threats To Validity}
\label{sec:threats}
This section analyzes the threats to the validity of our proposal.

\textbf{Internal Threats.} The major internal threat comes from the quality of the dataset created by Herzig \etal \cite{herzig2013s} and used in this paper. Labeling errors resulting from this manual process could affect classifier training and \textit{in fine} prediction quality. However, as a standard dataset, it is considered of good quality by many academics  \cite{terdchanakul2017bug,chawla2015automated,pingclasai2013classifying,luaphol2019feature,pandey2017automated,qin2018classifying} that work issue ticket classification. An existing bias would affect equally all state-of-the-art works based on it.

\textbf{External Threats.} 
External validity refers to the generalizibility of the treatment/condition outcomes.
Projects from which the dataset is generated are open-source and only written in Java. If we have a close look to selected features some are Java-dependant: class names (\eg \texttt{NullPointerExpection}, \texttt{DefaultHttpClient}, \texttt{FileInputStream}) or Java keywords (\eg \texttt{new}, \texttt{catch}, \texttt{throws}) that are used in ticket description are selected as features. Thus, generalization to projects written with other programming languages and technologies is yet to be studied.

\noindent The performance of AI techniques is also sensitive to computation resources. Comparing resource consumption for different approaches is tricky because of the diverse hardware and software used. Our classifier training process, while rather complex, was executed on a classic PC configuration (Intel core i5-7600 cadenced to 3.5 GHz with 8GB of RAM) and takes from a few minutes (training of a single MLP) to a few days (training and evolution of a large population of MLPs). However, resource consumption is not an actual limitation for this approach as it only regards the evolutionary selection and the training of the MPLs. Once selected and trained, the use of the MLP is fast and has a small footprint. Moreover, best results were obtained with the smaller MLP populations.  

%% file: Sections/Related_works.tex
\section{Related Works}
\label{sec:Related_works}
Ticket classification is a widely studied field.  If we extend the scope of related works beyond the binary bug classification solutions presented and discussed in \secref{sec:Background}, we see that many statistical models have been used for closely related problems. 

Few approaches exist that perform bug ticket classification on alternate datasets. 
Kallis \etal \cite{KALLIS2021102598} propose a binary bug ticket classifier based on a pre-trained text classification tool and transfer learning. In a following article \cite{KallisSCP19}, they develop a \github plugin named Ticket Tagger that automatically classifies tickets as they are written (on the fly).  As compared to our proposal, they use a wider yet not curated dataset. Indeed, they have collected 30,000 tickets from 12,112 heterogeneous projects labbeled in an \textit{ad hoc} manner by project contributors. Unlike the dataset proposed by Herzig \etal \cite{herzig2013s}, this one has not been manually curated and can thus supposedly suffer form flaws. A perspective is to compare our solution to theirs.

Other closely related approach perform multi-class ticket classification (\ie a more precise ticket classification in more than two predefined categories), 
bug-assignment (\ie automatic distribution of bug-tickets to expert developers), bug severity evaluation and duplicated bug ticket  detection. Some of these approaches have commonalities with the proposal of this paper:
\begin{itemize}
\item \textbf{Some use neural networks as their classification technique.} Kukkar \etal \cite{kukkar2019novel}, for example, have worked on bug severity classification. They have developed a method to classify bug ticket severity based on deep learning with convolutional neural networks mixed with random forest boosting.

\item \textbf{Some use the same labeled ticket dataset.}  Limsettho \etal \cite{LimsetthoHM14} have proposed a multi-class classification approach, based on Herzig \etal  dataset \cite{herzig2013s}, in order to dispatch tickets into five categories: Request For Enhancement, Bug, Improvement, Task and Test. They used NLP techniques combined with Hierarchical Dirichlet Process and Latent Dirichlet Allocation.


\item \textbf{Some use genetic algorithms to fine-tune their classifiers.} For example, Miller \etal \cite{miller1989designing} and Whitley \etal \cite{whitley1990genetic} have used GAs to find an optimized solution for connecting neurons. Another way of optimizing neural networks is to use GAs for the improvement of feed forward neural network weights \cite{572107, whitley1990genetic}. Others studies \cite{1176129, bashiri2011tuning} also fine-tuned hyper-parameters using GAs.
\end{itemize}

\noindent Even if the problems they tackle are slightly different from ours, these related work inspired us and comforted our choices and results.

%% file: Sections/Conclusion.tex
\section{Conclusion}
\label{sec:conclusion}

In this paper, we propose a solution for the automatic classification of issue tickets into two categories: bug related or not. Our solution combines natural language processing (TF-IDF), machine learning (MLP) and optimisation (genetic algorithms). For each technique we use, we have extensively studied the influence of their main options and parameters on classification performance. The hyper-parameters of the MLP are ultimately automatically optimized thanks to the GA we designed and implemented. Our solution has been validated on a standard dataset provided by Herzig \etal containing 5,991 tickets coming from 3 popular software projects. Our results can therefore be rigorously compared with state-of-the-art works. We computed F1 measure, recall and precision for our best MLP classifier, optimized with our genetic algorithm. We obtained a F1 Measure of 0.896 on cross-project tickets (\ie whole dataset). This corresponds to a noticeable performance improvement (gain of +0.082) as compared the best score established by Terdchanakul \etal on the same dataset. The balance between recall (0.881) and precision (0.913) is reasonably good, confirming the quality of our classifier.

\noindent These results open many perspectives. A first idea is to further improve classifier quality by optimizing more parameters with the genetic algorithm. In the same way, we also plan to study if the performance of the genetic algorithm itself can be further improved. Another perspective is to evaluate the robustness and genericity of this kind of approach, when applied to larger sets, possibly mixing different programming languages and technologies. Testing the performance of our solution on other ticket classification problems would also be very interesting, as ticket binary classification for other kinds of issues or multi-class classification such as bug sub-category prediction. A practical perspective is to experiment the use of automatic classification as an assistant during ticket redaction by developers. As the performance of automatic classification is high (at least for bugs), a misclassification could indicate an unclear content to be improved. First experiments on a prototype bug ticket writer assistant have been made and are promising. A first version of such an assistant is available online.\footnote{\url{http://ec2co-ecsel-1iegdism3qjis-2023048106.eu-west-3.elb.amazonaws.com/}}
